\documentclass[runningheads]{llncs}
\usepackage[T1]{fontenc}

\usepackage{indentfirst}
\usepackage{graphicx}
\usepackage{amssymb}
\usepackage{multirow}
\usepackage{threeparttable}

\begin{document}

% \title{Contrastive Knowledge Distillation for Robust Multimodal Learning}
\title{Contrastive Knowledge Distillation\\ for Robust Multimodal Sentiment Analysis}

\author{Zhongyi SANG\inst{1} \and
Kotaro FUNAKOSHI\inst{1} \and
Manabu OKUMURA\inst{1}}
\authorrunning{Z. SANG et al.}
\institute{Tokyo Institute of Technology, Tokyo, Japan}

% \author{Anonymous authors}
% \institute{Anonymous addresses}

\maketitle              % typeset the header of the contribution
\begin{abstract}
Multimodal sentiment analysis (MSA) systems leverage information from different modalities to predict human sentiment intensities. Incomplete 
% data 
modality is an important issue that may cause a significant performance drop in MSA systems. By generative imputation, i.e.,  recovering the missing data from available data, systems may achieve robust performance but will lead to high computational costs. 
This paper introduces a knowledge distillation method, called `Multi-Modal Contrastive Knowledge Distillation' (MM-CKD), to address the issue of incomplete modality in video sentiment analysis with lower computation cost, as a novel non-imputation-based method. 
We employ Multi-view Supervised Contrastive Learning (MVSC) to transfer knowledge from a teacher model to student models. 
This approach not only leverages cross-modal knowledge but also introduces cross-sample knowledge with supervision, jointly improving the performance of both teacher and student models through online learning. Our method gives competitive results with significantly lower computational costs than state-of-the-art imputation-based methods.

\end{abstract}

\section{Introduction}
In recent years, with the rapid growth of various social media platforms, people share their lives, ideas, or interesting experiences on various platforms. 
People express themselves through multimodal ways such as words, speech, and facial expressions, there has been a growing interest in leveraging multimodal data from online platforms~\cite{chandrasekaran2021multimodal}.

Sentiment Analysis (SA) is a task that aims to extract opinions or sentiment intensities from people's expressions. 
Since last century, SA has been an important research field of natural language processing (NLP)~\cite{liu2022sentiment,majumder2019dialoguernn}. 
However,  human expression is a complex process~\cite{mehrabian1971silent}, not only lexical information, but also voice intonation and facial expressions can carry information and help in problems like ambiguity, humor, and sarcasm.

Researchers have been incorporating multimodal information into the SA task in recent years. 
Multimodal Sentiment Analysis (MSA) systems can leverage all available information during communication to predict sentiment intensities. 
To combine information from diverse modalities, various feature fusion methods have been proposed, such as tensor fusion~\cite{zadeh2017tensor}, graph fusion~\cite{zadeh2018multimodal}, and so on. 
In recent years, researchers have found Transformer~\cite{vaswani2017attention} a potential fusion method in multimodal learning~\cite{ma2023multimodal,wang2024incomplete}. 
However, in real-world scenarios, we cannot ignore the possibility of missing modalities, 
which leads to a lack of robustness of the MSA system~\cite{ma2022multimodal}. 
By generative \textit{imputation}, i.e.,
recovering the missing data from available data, MSA systems may achieve robust performance but will lead to high computational costs.
%Incomplete multimodal learning is a subtask of multimodal learning, which enables MSA systems to be able to cope with incomplete subsets of multimodal information. 

In this paper, we propose the Multi-Modal Contrastive Knowledge Distillation (MM-CKD) method to transfer information from a complete multimodal teacher model to student models for incomplete modalities, which can avoid the additional cost of recovery and realize a competitive performance, as a \textit{non-imputation-based} method.
With the help of a strong teacher model, robust student models are trained to learn representations of subsets of modalities. 
We propose using a Multiview Supervised Contrastive Learning (MVSC)~\cite{khosla2020supervised} for representative feature learning and information transfer between the teacher and student models at the same time, which can improve the performance of the teacher model and student models simultaneously.
The contributions of our work can be summarized as:

(1) We propose MM-CKD, a non-imputation, multimodal contrastive knowledge distillation method on incomplete MSA, i.e., MSA with missing modalities.

(2) MM-CKD is much lighter than state-of-the-art (SOTA) imputation-based method~\cite{hazarika2022analyzing,wang2024incomplete}.

(3) Our experiment results on CMU-MOSI~\cite{zadeh2016mosi} and CMU-MOSEI~\cite{zadeh2018multimodal} prove the effectiveness of our method as a SOTA non-imputation-based method.

\section{Related Work}\label{ch:background}

\subsection{Incomplete Multimodal Learning}

We refer to the problem of learning a multimodal model that can handle missing modalities in input as \textit{incomplete multimodal learning}. 
Lian et al.~\cite{lian2023gcnet} summarized existing approaches to incomplete multimodal learning by categorizing them into imputation-based and non-imputation-based methods. 
The key distinction is that imputation-based methods recover the missing data as a part of the complete input, however, non-imputation-based methods do not.

\textbf{Imputation-based methods} can be classified by the methods used to recover the missing data. 
Intuitively, zero or average values can pad the incomplete input~\cite{parthasarathy2020training}. 
However, zero padding ignores the missing supervised information and only uses zero values for missing information which will remain the gap between recovered data and original distribution. 
Average padding uses the average value within the same class, which can introduce label-related information when recovering.
Conventional DNN-based methods use a generative decoder on the output of the available encoder to recover the missing modalities' information. 
Cai et al.~\cite{cai2018deep} used such a framework to reconstruct the PET modality with MRI for medical diagnosis. 
Joint representation learning is another direction in DNN-based methods.
Pham et al.~\cite{pham2019found} proposed a sequence-to-sequence (seq2seq) architecture called MCTN to extract joint representation when transferring one modality to another target modality. 
They observed that the encoder's output serves as a joint representation of the two modalities, with additional information being recovered in the decoder.

As more attention is paid to this field, researchers realize the gap between recovered data and original distribution can significantly affect performance. 
Wang et al.~\cite{wang2024incomplete} introduced a distribution-consistent modality recovering method with a score-based diffusion model. 
The diffusion model can transfer random noise into the distribution space of the missing modalities, which realizes the current SOTA performance on multimodal sentiment analysis benchmarks.

\textbf{Non-imputation-based methods}
do not recover the missing modalities but find other strategies to use available information better.

The grouping strategy is separating missing data into different groups based on missing patterns and conducting feature learning independently on different missing groups for the best performance on each of them. 
Li et al.~\cite{li2018multi} worked on multi-source block-wise missing data for Alzheimer’s disease. 
They use a series of problems for the original problem to realize a multi-task framework in group strategy.

Knowledge distillation, which was originally proposed as a model compression method~\cite{hinton2015distilling}, can also help with incomplete multimodal learning. 
With the help of knowledge distillation, information can be transferred between modalities. 
As an application in incomplete multimodal learning, Liu et al.~\cite{liu2023emotionkd} proposed a lightweight student model as the unimodal model. 
In Liu's case, such a student model is similar to the grouping strategy, which is used on some target missing groups, but the difference is the student model can learn from the teacher to realize information communication between different groups.

Without recovery, another efficient approach to cope with incomplete multimodal learning is maximizing correlations between different modalities.
These methods ensure different modalities of the same sample to have related representations. 
CCA~\cite{hotelling1992relations} learned relationships between multi-modalities by linearly mapping them into a low-dimensional space with maximal canonical correlations.
With the help of deep neural networks, Andrew et al.~\cite{wang2015deep} proposed DCCAE, which can learn more complex nonlinear information between modalities.

\subsection{Contrastive Learning}
After the success of CLIP~\cite{radford2021learning}, contrastive learning in multimodal learning has been widely used across different tasks~\cite{radford2021learning,wang2023cross,ma2023multimodal}. 

Chopra et al.~\cite{chopra2005learning} first found a simple distance (such as Euclidean distance) of low-dimensional output can approximate the neighborhood relationships in the high-dimensional input space.
Chen et al.~\cite{chen2020simple} proposed SimCLR with NT-Xent loss using cosine similarity without the reduction of dimensions. 

The strategy for splitting positive and negative samples can also introduce valuable information for contrastive learning.
In self-supervised learning, as demonstrated by CLIP~\cite{radford2021learning}, positive pairs are usually generated by data augmentation or using different views like binary and color images. 
Khosla et al.~\cite{khosla2020supervised} proposed the Supervised Contrative loss as an extension of N-pair loss~\cite{sohn2016improved} by leveraging labels to split samples.
For better performance on regression tasks, Ma et al.~\cite{ma2023multimodal} used the distance between regression labels to distinguish positive and negative samples. 

\subsection{Knowledge Distillation}

Knowledge Distillation (KD) was initially proposed as a model compression method, which can transfer knowledge in a large model (teacher) to a smaller model (student) without a significant loss in performance~\cite{hinton2015distilling}.
This technique has shown effectiveness in multimodal learning~\cite{liu2023emotionkd} by realizing cross-modal information communication.

KD can be divided into online and offline learning. 
Offline learning, a simple and easy method to realize knowledge distillation, was used in %Hinton's work~
\cite{hinton2015distilling} initially with a pre-defined teacher model.
However, online learning enables dynamic interaction between models by updating teacher and student models simultaneously.
In Deep Mutual Learning (DML)~\cite{zhang2018deep}, an ensemble of students is defined and teaches each other during training. 
Without a pre-defined powerful teacher, these students can still achieve compelling results on different tasks.

On the other hand, the performance of knowledge distillation varies with the type of knowledge transferred.
In the vanilla Knowledge Distillation~\cite{hinton2015distilling}, response-based knowledge is known as soft targets, like the softmax output in the classification task.
Feature-based knowledge is another popular knowledge to transfer, by using the output of intermediate layers in DNNs. 
SemCDK~\cite{wang2022semckd} is a cross-layer knowledge distillation, which transfers feature-based knowledge between the output of intermediate layers in the teacher and student model by attention-based Mean Squared Error (MSE).
Mixing multi-level knowledge~\cite{zhou2019m2kd} also has been shown to produce interesting effects in knowledge distillation.

\section{Proposed Method}\label{ch:proposedsytem}
A multimodal model takes input from different modalities and fuses them to fulfill a classification or regression task.
The performance of such a model is potentially sensitive to incomplete modality caused by data corruption and signal loss. 
We aim to cope with this incomplete multimodal learning problem.

\subsection{Task Definition}\label{sec:taskdef}
We have a complete MSA dataset $D = \{(X_i, y_i)\ \mid i \in \{1, ..., N\}\}$, where each sample $X_i$ with label $y_i$ consists of three modalities, i.e., $X_i= \lbrace X_i^l, X_i^v, X_i^a\rbrace $, for lexical ($l$), visual ($v$), and acoustic ($a$) information. 
$X_i^m$ ($m \in \{l,v,a\})$ is a $n$-length sequence of $d^m$-dimensional feature vectors, i.e., $X_i^m = [m^i_1, m^i_2, ..., m^i_n]$. 
Here, $n$ is not a constant and varies for each sample.
The label $y_i$ is a numerical value in a fixed range, and thus we handle a regression task from $X_i$ to $y_i$.
Hereafter, we omit data sample index $i$ unless necessary.

We consider the possible loss of information in modality-wise manners.
That is, possible input $X$ to a regression model would be reduced into one of 
the following incomplete modality cases:
$\lbrace X^l, X^v \rbrace$, $\lbrace X^v, X^a \rbrace $, $\lbrace X^l, X^a \rbrace $, $\lbrace X^l\rbrace $, $\lbrace X^a\rbrace $, and $\lbrace X^v\rbrace $. 
The model must predict the label $y$ even with incomplete modality.

\subsection{Overview}
The overview of our MM-CKD method is shown in Figure 1. 
MM-CKD consists of a teacher model with complete multimodal input and joint 6 student models on subsets of modalities. 
Each sub-model is designed as a conventional MSA model, with three function blocks:
extraction, fusion, and regression.

\begin{figure}[t]
\includegraphics[width=\textwidth
]{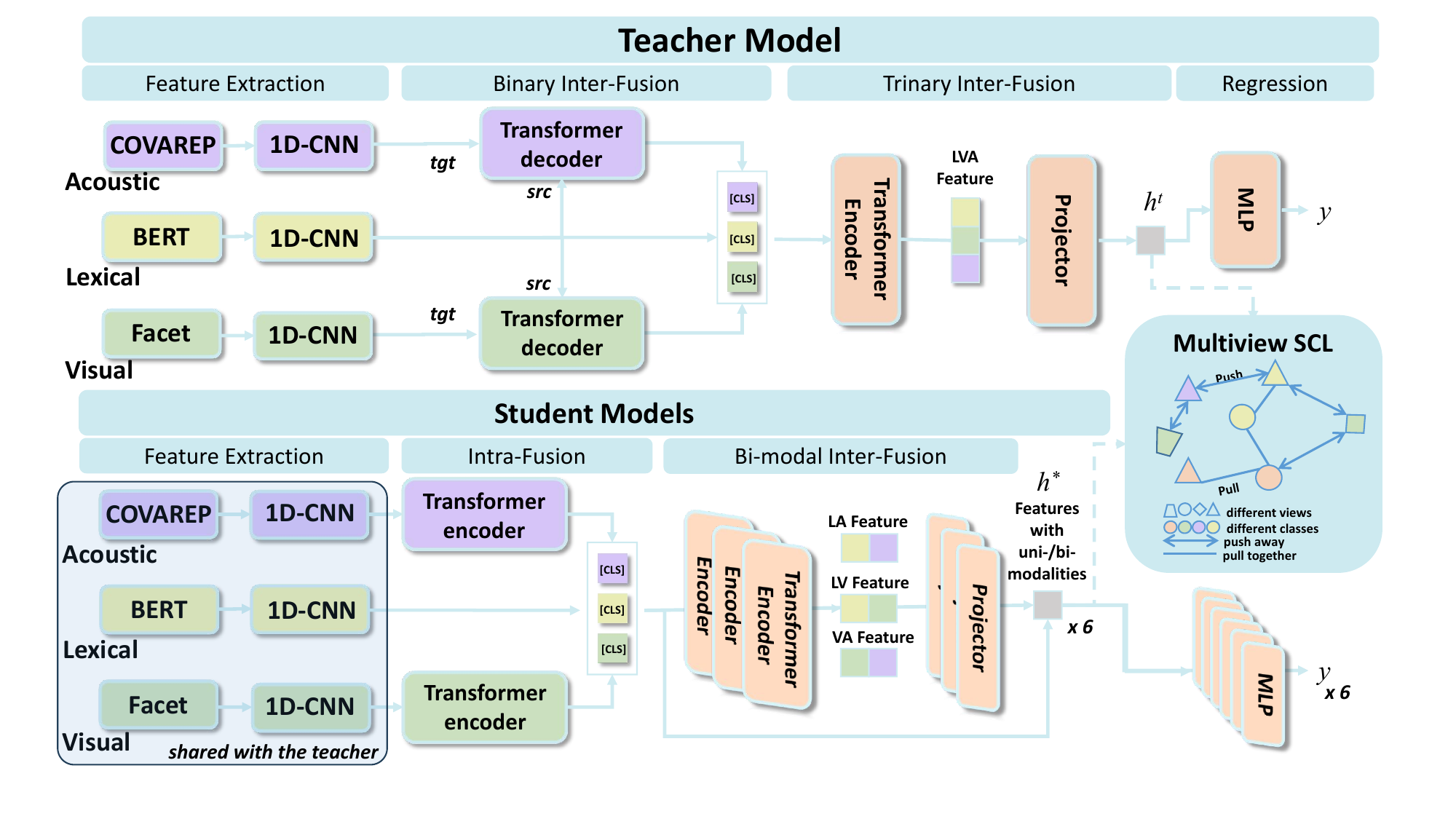}
\caption{The overview of the proposed Multi-Modal Contrastive Knowledge Distillation} \label{fig1}
\end{figure}

\subsection{Teacher Model}

\subsubsection{Feature Extraction}\label{sec:fea_extract}
$X^l$ is a contextual embedding sequence processed by BERT~\cite{devlin2018bert} from a transcribed utterance, which includes a special [CLS] token at the head. 
$X^v$ and $X^a$ are feature vector sequences obtained by Facet~\cite{ekman1997face} and COVAREP~\cite{degottex2014covarep}, respectively. 
The input data from different modalities with a length $n$ are represented as $X^m \in \mathbb{R}^{d^m \times n}$. 
Then these sequences of representations are converted into $\bar{X}^m \in \mathbb{R}^{d \times n}$ of the uniform dimension $d$ using 1D-CNN. 

\subsubsection{Feature Fusion Using Dot-Product Attention}

Feature fusion in MM-CKD is realized by Transformer~\cite{vaswani2017attention}. 
Our model leverages three different Transformer encoders/decoders for feature fusion. We utilize encoders with self-attention to fuse sequential intra-modal information and the stacked inter-modal utterance-level features. Decoders with cross-modal attention can augment acoustic and visual modality with the lexical sequence in our teacher model.

\subsubsection{Binary Inter-Modal Feature Fusion}
As the first step of our inter-modal fusion, we use a Transformer decoder to fuse sequences from two different modalities. 
As lexical information plays the most important role among the three modalities in MSA benchmarks~\cite{hazarika2022analyzing},
we first fuse lexical modality and the other two modalities following previous work~\cite{ma2023multimodal}. 
We use $\bar{X}^l$ as source input (used as key/value of dot-product cross attention) 
and $\bar{X}^m \ (m \in \{v, a\})$ as target input (used as a query) of Transformer decoder $decoder_m(\cdot)$ for modality $m$ is defined as below:

\begin{equation}
D^m = decoder_m(\bar{X}^l, [m_\mathrm{cls}] + \bar{X}^m).
\label{eq:decoder_m}
\end{equation}

$m_\mathrm{cls}$ is a learnable embedding vector corresponding to a [CLS] token for modality $m$.
$decoder_m$ uses the originally proposed sinusoidal positional embedding~\cite{vaswani2017attention} on the target input. 

Note that we do not use causal masking, which is used in common auto-regressive generation with the Transformer decoder. 
We believe this can help the learnable embedding vector to capture more information.
The cross-attention in the decoder is intended to decrease the gap between the three modalities by augmenting the representation of two secondary modalities with the primary lexical modality and realizing the first step of inter-modal feature fusion. $D^m \ (m \in \{v,a\})$ are augmented sequence of visual and acoustic features.

\subsubsection{Trinary Inter-Modal Feature Fusion}
As the second step of our inter-modal fusion,
we integrate $\bar{X}^l$, $D^v$, and $D^a$ from the previous stages.
For this integration, we only use the first elements of the three sequences, which correspond to [CLS] tokens.
We stack them into one matrix $F = [f_l, f_a, f_v]$, where $f_l = \bar{X}^l[0]$, $f_a = D^a[0]$, and $f_v = D^v[0]$. Here $[0]$ indicates the head element of a sequence.
Then, we process $F$ with a transformer encoder $encoder_f(\cdot)$ defined as below:

\begin{equation}
[\hat{f_l}, \hat{f_a}, \hat{f_v}] = encoder_f(F).
\label{eq:encoder_f}
\end{equation}

We concatenate the output of $encoder_f$ as one vector
$\hat{f} = \hat{f}_l\oplus\hat{f}_a\oplus\hat{f}_v \in \mathbb{R}^{3d}$ 
and feed it into a learnable projector with a single linear layer to get the final fused feature $h_t \in \mathbb{R}^{d}$ of the teacher model as:
$h^t = projector(\hat{f}).$

\subsection{Student Model}
The main difference between our and existing methods is that MM-CKD uses knowledge distillation from a teacher model to student models to resolve incomplete multimodal learning in MSA.
The student models are introduced to learn features from incomplete subsets of modalities with the knowledge learned by the teacher model mentioned above.

\subsubsection{Feature Extraction}
The feature extraction in the student models is shared will the teacher model to save computation resources and keep knowledge learned by the teacher model in feature extraction.

\subsubsection{Intra-Modal Feature Fusion}
The binary inter-modal feature fusion in the teacher model is replaced by the intra-modal feature fusion in the student models because of the incomplete modalities in input.
Intra-modal feature fusion is performed with two encoders $encoder^s_m$ only for the visual and acoustic modalities.
That is, 
\begin{equation}
S^m = encoder^s_m([m_\mathrm{cls}] + \bar{X}^m),
\end{equation}
where $m \in \{v,a\}$. 

Same as in the teacher model, we introduce a learnable embedding vector $[m_\mathrm{cls}]$ corresponding to a [CLS] token for modality $m$.
Two $encoder^s_m$ also use the sinusoidal positional embedding~\cite{vaswani2017attention} on the sequential input.
For lexical information, we reuse $\bar{X}^l$ to save computation resource, i.e, $S^l = \bar{X}^l$. 

\subsubsection{Bi-Modal Feature Fusion}
To cope with the bi-modal incomplete input cases mentioned in Section~\ref{sec:taskdef} (i.e., 
$\lbrace X^l, X^v \rbrace$, $\lbrace X^v, X^a \rbrace $, $\lbrace X^l, X^a \rbrace $), we build three feature representations by fusing two of the three modalities.

For this purpose, first, we employ the same approach as the trinary inter-modal feature fusion (Eq.~\ref{eq:encoder_f}) for the teacher model.
That is, 
\begin{equation}
[\hat{f}_{m_1}, \hat{f}_{m_2}] = encoder^s_{m_1,m_2}(F^s_{m_1m_2}),
\end{equation}
where the pair of 
$m_1$ and $m_2$ is any of the three ordered combinations of $[l,v]$, $[l,a]$, or $[a,v]$, and $F^s_{m_1m_2} = [S^{m_1}[0], S^{m_2}[0]]$. 

Then, we concatenate the output 
as one vector $\hat{f}^s_{m_1m_2} = \hat{f}^s_{m_1}\oplus\hat{f}^s_{m_2} \in \mathbb{R}^{2d}$ 
and project it by a learnable projector with a single linear layer to get the final bi-modal feature representations $h^{m_1m_2} \in \mathbb{R}^{d}$  in the student model as below:
\begin{equation}
h^{m_1m_2} = projector^s_{m_1m_2}(\hat{f}^s_{m_1m_2}).
\end{equation}
Uni-modal feature representations $h^m \ (m \in \{l,v,a\})$ are defined as $h^m = S^{m}[0]$.

\subsection{Contrastive Learning and Model Training}
\subsubsection{Multi-view Contrastive Learning for a Regression Task} 
We employ the multi-view supervised contrastive learning~\cite{khosla2020supervised}. 
Here, ``multi-view'' refers to augmented samples for one data sample. 
In this work, the incomplete modality cases in Section~\ref{sec:taskdef} are the different views of the complete data $X = \{X^l, X^v, X^a\}$.
While supervised contrastive learning uses labels to divide the positive and negative samples for a classification task, the multi-view loss (Eq.~\ref{eq:mvsc}) treats different views of one sample as positive samples each other in addition to the samples of the same class with the sample in question.

On the other hand, our task is a regression task to predict the numerical value $y_i$. 
Thus, we use a distance-aware contrastive loss for a regression task~\cite{ma2023multimodal}. 
For a batch $B$ of complete data samples $\{X_i\}_i$, an augmented set of feature representations $V$ is prepared as
$V = \bigcup_{i} \{h^t_i, h^{la}_i, h^{lv}_i, h^{av}_i, h^l_i, h^a_i, h^v_i \}$.
Thus, $|V| = 7|B|$.
Then, for each sample $h_j$ in $V$, absolute distance $| y_p - y_j |$ is used to define a set of positive samples $P(j) \subseteq V$ with a certain threshold $\lambda$.
That is, 
$P(j)=\{h_p \in V \mid | y_p - y_j | \leq \lambda \}$.
Our MVSC loss is defined as:
\begin{equation}
L_{MVSC} = \sum_{h_j \in V} \frac{-1}{\left| P(j) \right|} \sum_{h_p \in P(j)} log \frac{exp(h_j \cdot h_p / \tau)}{\sum_{h_a \in V \setminus \{h_j\}} exp(h_j \cdot h_a / \tau)}.\label{eq:mvsc}
\end{equation}

\subsubsection{Training}
We employ 7 different two-layer MLPs with the same hidden dimension $d_{hid}$ as regression heads to make different predictions with the augmented set of feature representations $V$. 
The definition of each two-layer MLP is as below:

\begin{equation}
H_h = \sigma(W_1h+B_1), \ \
MLP_h(h) = W_2H_h+B_2,
\end{equation}
where $W_1,W_2$ are weights, and $B_1,B_2$ are biases for two layers. 
$H_h \in \mathbb{R}^{d_{hid}}$ is the hidden variable in MLP. $\sigma$ is the activation function on the first layer, which is $ReLU$ in our method.
For the teacher model, the prediction on complete feature representation $h^t$ is made by the MLP as: 
$\hat{y}_t = MLP_t(h^t)$.

Different from the teacher model, there are 6 different predictions made by the student model to solve 6 incomplete multimodal problems. 
The regression heads in the student model are defined as:
$\hat{y}_m = MLP_t(h^m)$,
where $m \in \{l,v, a, la, lv, av\}$.

We employ online learning for knowledge distillation. 
The teacher and student models are updated together as a joint optimization problem.
We use mean absolute error (MAE) to represent the gap between model prediction $\hat{y}$ and the ground truth $y$. 
The joint optimization loss for regression tasks is as below:

\begin{equation}
L_\mathrm{regression} = MAE(\hat{y}_t,y) + \sum_{m \in \{l, v, a, la, lv, av\}} MAE(\hat{y}_m,y),
\label{eq:regression}
\end{equation}
where the first part represents the loss for training the teacher model. 
The second part represents the joint loss for training the student models.

Another important part of the training target is contrastive learning. 
We calculate the MVSC loss in Eq.~\ref{eq:mvsc} for contrastive learning. 
The overall learning of our proposed model is performed by minimizing:
$L_c = L_\mathrm{regression} + L_{MVSC}$.

\section{Evaluation}\label{ch:evaluation}
\subsection{Datasets}
\textbf{CMU-MOSI}~\cite{zadeh2016mosi}
is the first opinion-level corpus for multi-modal sentiment analysis. 
It has been a popular benchmark in MSA tasks since collected in 2016. 
The dataset consists of 93 randomly selected videos featuring 89 different speakers from YouTube.
These videos are separated into 2,199 utterance-video segments with subjective sentiment annotations. 
Sentiment is annotated as a value of intensity ranging from -3 to 3, %representing 
i.e., strongly negative to strongly positive.

\textbf{CMU-MOSEI}~\cite{zadeh2018multimodal} is an updated version of CMU-MOSI. 
Much larger than the previous CMU-MOSI, this dataset contains 23,453 annotated video clips from 5,000 videos, 1,000 distinct speakers, and 250 topics. 
Sentiment is annotated in the same way as CMU-MOSI.

\subsection{Evaluation Metrics and Implementation Details}
\subsubsection{Evaluation Metrics}
For both datasets, the sentiment intensity score ranges from -3 to 3, representing strongly negative, negative, weakly negative, neutral, weakly positive, positive, and strongly positive. 
We calculate two evaluation metrics to compare our method with existing works: binary accuracy ($Acc_2$) and seven-class accuracy ($Acc_7$). 
In detail, the binary accuracy is calculated on positive and negative samples, and samples with neutral intensity will be excluded. 
Seven-class accuracy is calculated on all seven classes of intensities and reflects the performance of our method on each type of sentiment intensity.

\subsubsection{Implementation Details}
Following the existing works, we evaluate the robustness of our method on two different missing protocols by making a fair comparison with baseline methods. 

In the \textit{fixed missing protocol}, each sample's missing modality is consistent.
As our datasets have three modalities, one or two modalities will be completely removed. 
Results on six subsets will be reported under this protocol and the input subsets of modalities are $m \in \{ l, a, v, la, lv, av \}$. 

The second protocol is the \textit{random missing protocol}. 
In this protocol, the missing modalities will be random for each sample. 
The missing rate (MR) is defined as $MR = 1 - \frac{1}{L \times M}{\sum_{i=1}^{L} a_i}$, where $L$ is the number of samples, $a_i$ denotes the number of available modalities in $i^{th}$ sample and $M$ is the number of modalities which equals $3$ here. 
To ensure there is no empty input, we keep at least one modality left in each sample, which leads to the missing rate (MR) being smaller than 0.7 in this protocol. 
We use the sampling method proposed by Wang et al.~\cite{wang2023distribution} to make a fair comparison in the following parts. 

We implemented all the experiments using PyTorch on an RTX 3090 GPU with 24GB memory. 
We set the training batch size as 64 for both CMU-MOSI and CMU-MOSEI. 
The learning rate is set to $lr=1e-4$ for CMU-MOSI and  $lr=5e-5$ for CMU-MOSEI.
The input dimensions are: $d^{l} = 768$, $d^{a} = 74$ and $d^{v} = 47$ for CMU-MOSI, $d^{v} = 35$ for CMU-MOSEI. 
The hidden size of Transformer encoders and decoders are defined as $d_{model} = 32$. 
Multihead attention in our Transformer encoders and decoders is defined with $head\_num = 4$.
The depth of our Transformer encoders and decoders is set to $N = 2$.
At last, the hidden size of MLPs as regression heads is $d_{hid} = 128$.
Another two hyper-parameters in Eq.~\ref{eq:mvsc} are set as $\lambda = 0.9$ and $\tau = 0.1$.

\subsection{Results}
We compare MM-CKD with non-imputation-based and imputation-based SOTA methods. 
{DCCAE}~\cite{wang2015deep}
is a non-imputation-based method, 
{DicMOR}~\cite{wang2023distribution}, and {IMDer}~\cite{wang2024incomplete} are imputation-based methods. 
In the following tables, results of MMCKD are average values from five random seeds. 
Results of DCCAE, DicMOR and IMDer are reported values in their publications~\cite{wang2023distribution,wang2024incomplete}.
Table~\ref{tab1} shows the result for the fixed missing protocol.
MM-CKD achieves a better result than all Non-imputation-based works and a better result than most imputation-based-method except {IMDer}. 
Table~\ref{tab2} shows the result for random missing protocol, our method also beats most SOTA methods and achieves the best performance with complete input, i.e., $MR = 0.0$, on CMU-MOSI,
and for $Acc_2$ on CMU-MOSEI . 

\begin{table}[t]
\centering
\small
\caption{Fixed missing protocol.}\label{tab1}
\begin{threeparttable}
\tabcolsep=0.9mm
\begin{tabular}{c|c|c|c|c|c|c}
\hline
\multicolumn{7}{c}{CMU-MOSI}\\
\hline
Model & L+A & L+V & A+V & L & V & A\\
\hline
DCCAE~\cite{wang2015deep} & 77.0/30.2 & 76.7/30.0 & 54.0/17.4 & 76.4/28.3 & 52.6/17.1 & 48.8/16.9 \\
DicMOR~\cite{wang2023distribution} & \underline{85.5}/\underline{45.2} & 85.5/44.6 & \underline{64.0}/\underline{21.9} & 84.5/44.3 & \underline{62.2}/20.9 & \underline{60.5}/20.9\\
IMDer~\cite{wang2024incomplete} & 85.4/45.0 & \underline{85.5}/\underline{45.3} & \textbf{63.6}/\textbf{23.8} & \underline{84.8}/\underline{44.8} & \textbf{61.3}/\textbf{22.2} & \textbf{62.0}/\textbf{22.0}\\
\hline
MM-CKD & \textbf{86.6}/\textbf{46.5} & \textbf{86.9}/\textbf{46.2} & 56.9/21.6& \textbf{86.5}/\textbf{46.4} & 56.5/\underline{20.9} & 56.1/\underline{21.1}\\
\hline
\multicolumn{7}{c}{CMU-MOSEI}\\
\hline
Model &  L+A & L+V & A+V & L & V & A\\
\hline
DCCAE~\cite{wang2015deep} & 80.0/47.4 & 80.4/47.1 & 62.7/41.6 & 79.7/47.0 & 61.1/40.1 & 61.4/40.9\\
DicMOR~\cite{wang2023distribution} & 84.3/51.3 & 84.3/51.1 & \underline{64.1}/42.0 & 84.2/\underline{52.4} & \underline{63.6}/\underline{42.0} & \underline{62.9}/41.4\\
IMDer~\cite{wang2024incomplete} & \underline{85.1}/\textbf{53.1}&  \underline{85.0}/\textbf{53.1}& \textbf{64.9}/\textbf{42.8}& \underline{84.5}/\textbf{52.5}& \textbf{63.9}/\textbf{42.6}& \textbf{63.8}/\textbf{41.7}\\
\hline
MM-CKD & \textbf{87.8}/\underline{51.7} & \textbf{88.1}/\underline{52.0} & 62.5/\underline{42.0} & \textbf{88.4}/51.5 & 62.6/41.9& 62.6/\underline{41.4}\\
\hline
\end{tabular}
\end{threeparttable}
\\
Values in each cell denote $Acc_2$/$Acc_7$. \textbf{Bold} is the best and \underline{underline} is the second.

\vspace{0.5cm}
\caption{Random missing protocol.}\label{tab2}
\begin{threeparttable}
\tabcolsep=1.2mm
\begin{tabular}{c|c|c|c|c|c}
\hline
\multicolumn{6}{c}{CMU-MOSI}\\
\hline
Model & 0.0 & 0.2 & 0.4 & 0.6 & Avg.\\
\hline
DCCAE~\cite{wang2015deep} & 77.3/31.2 & 71.8/27.6 & 63.6/24.2 & 59.6/20.9 & 68.1/26.0 \\
DicMOR~\cite{wang2023distribution} & 85.7/45.3 & 82.1/42.3 & \underline{77.9}/37.6 & \underline{73.3}/32.7 & \underline{79.8}/39.5\\
IMDer~\cite{wang2024incomplete} & \underline{85.7}/\underline{45.3}  & \textbf{83.5}/\textbf{44.3} & \textbf{78.6}/\textbf{39.7} & \textbf{74.7}/\textbf{35.8} & \textbf{80.6}/\textbf{41.3}\\
\hline
MM-CKD & \textbf{87.0}/\textbf{46.4} & \underline{82.2}/\underline{42.3} & 77.2/\underline{38.1} & 70.9/\underline{33.8} & 79.3/\underline{40.2}\\
\hline
\multicolumn{6}{c}{CMU-MOSEI}\\
\hline
Model & 0.0 & 0.2 & 0.4 & 0.6 & Avg.\\
\hline
DCCAE~\cite{wang2015deep} & 81.2/48.2 & 75.5/46.3 & 70.3/44.0 & 67.6/42.9 & 73.7/45.4 \\
DiCMoR~\cite{wang2023distribution} & 85.1/53.4 & \underline{81.8}/\underline{51.4} & \underline{78.7}/\underline{48.8} & \underline{76.7}/\underline{46.8} & \underline{80.6}/\underline{50.1}\\
IMDer~\cite{wang2024incomplete} & \underline{85.1}/\textbf{53.4} & \textbf{82.7}/\textbf{52.0} & \textbf{79.3}/\textbf{50.0} & \textbf{78.0}/\textbf{48.5} & \textbf{81.3}/\textbf{51.0}\\
\hline
MM-CKD & \textbf{88.3}/\underline{52.6} & 81.4/49.6 & 77.0/48.3 & 74.4/46.5 & 80.3/49.3\\
\hline
\end{tabular}
\end{threeparttable}
\\
Values in each cell denote $Acc_2$/$Acc_7$. \textbf{Bold} is the best and \underline{underline} is the second.
\end{table}

Imputation-based models often suffer a high cost in computation due to the optimization of the imputation loss. 
Table~\ref{tab3} shows the computation cost of two SOTA imputation-based methods compared to our MM-CKD method. 
We compare the training time of three methods on the CMU-MOSI dataset, which is split into a training set of 1284 samples and a test set of 686 samples. 
For a fair comparison, we align all models to the same batch size in the test comparison. 
The training cost is calculated on the reported batch size to ensure the best performance of each model.
Additionally, the FLOP is calculated on the same batch of test samples with $bs = 4$.
The result indicates MM-CKD significantly reduces the cost of computation resources, compared with two SOTA imputation-based methods, which have a performance better than or equal to ours in Table~\ref{tab1} and Table~\ref{tab2}.

\begin{table}[tb]
\centering
\small
\caption{Computation cost on CMU-MOSI.}\label{tab3}
\begin{threeparttable}
\tabcolsep=2mm
\begin{tabular}{c|c|c|c|c}
\hline
Model & Model size & Train(bs=reported) & Epochs & Test FLOPs\\
\hline
DiCMoR~\cite{wang2023distribution} & 431M & 67.4s (bs=4)  & 50 & $17.4\times10^9$\\
IMDer~\cite{wang2024incomplete} & 467M & 28.6s (bs=32) & 80 & $17.4\times10^9$\\
MM-CKD & 426M & 5.3s (bs=64) & 65 & $5.2\times10^9$\\
\hline
\end{tabular}
%Note: 
\end{threeparttable}
\\The experiments are conducted on the same RTX 2080Ti.
\end{table}

Talbe~\ref{tab4} shows
two ablation studies on CMU-MOSI under the fixed missing protocol to evaluate the effectiveness of the following two important components in MM-CKD. 

\paragraph{1. Contrastive Learning}
The performance of the teacher model using different contrastive learning methods was compared.
To prove the effectiveness of our MVSC, we compare it with the performance on uni-view supervised contrastive learning and the performance without contrastive learning.

\paragraph{2. Knowledge Distillation}
We compare the performance of incomplete multimodal learning in the student model in the second part of Table~\ref{tab4}.
MM-CKD is compared against the conventional mean squared error (MSE) loss for feature level knowledge distillation~\cite{liu2023emotionkd} and no knowledge distillation.
The MSE loss can be calculated in three different combinations to align different aspects of information with the teacher model. The following MSE losses will be used in $L_c$ %Eq.~\ref{eq:joint} 
to replace the $L_{MVSC}$ for the ablation study.

1. MSE(v/a): MSE is calculated over the output of intra-modal feature fusion ($h^v$/$h^a$) in the student models and output of binary intra-fusion ($f_v$/$f_a$) in the teacher model, here is:
\begin{equation}
L_{MSE(v/a)} = \sum_{m \in \{v, a\}} MSE(h^m, f_m). 
\end{equation}
2. MSE(la/lv/av): MSE is calculated over the output of bi-modal feature fusion ($h^{la}$/$h^{lv}$/$h^{av}$) in the student models and output of trinary inter-modal feature fusion ($h^t$) in the teacher model, here is:
\begin{equation}
L_{MSE(la/lv/av)} = \sum_{m \in \{lv, la, av\}} MSE(h^m, h^t). 
\end{equation}
3. MSE(all): The combination of two methods above, here is:
\begin{equation}
L_{MSE(all)} = L_{MSE(v/a)} + L_{MSE(la/lv/av)}. 
\end{equation}

Table~\ref{tab4} presents all results above to avoid unfair comparison.
The None set refers to training the student directly on the second part of $L_{regression}$ in Eq.~\ref{eq:regression}, without any joint optimization.
Results under fixed missing protocol are the average value on three bi-modal subsets: \{L+V, L+A, A+V\} and three uni-modal subsets: \{L, A, V\}.

From the quantitative result, we can conclude the effectiveness of our proposed method and the potential of knowledge distillation in robust MSA. 

\begin{table}[tb]
\centering
\small
\caption{Results of Ablation Study on CMU-MOSI.}\label{tab4}
\begin{threeparttable}
\tabcolsep=1.2mm
\begin{tabular}{c|c|c|c|c}
\hline
Component & Set & L+A+V & Avg. \{L+V, L+A, A+V\} & Avg. \{L, A, V\}\\
\hline
\multirow{3}{*}{CL} & Multi-view  & \textbf{87.0}/\textbf{46.4} & -- & --\\
~ & Uni-view & 86.8/46.4 & -- & -- \\
~ & None & 86.1/43.6 & -- & --\\
\hline
\multirow{5}{*}{KD} & Multi-view  & -- & \textbf{76.8}/\textbf{38.1} & 66.4/\textbf{29.5}\\
~ & MSE(all) & -- & 76.8/37.2 & 65.2/28.5 \\
~ & MSE(v/a) & -- & 77.2/37.0 & \textbf{67.2}/28.8\\
~ & MSE(la/lv/av) & -- & 77.2/37.0 & 67.0/28.6\\
~ & None & -- & 75.1/35.6 & 64.2/28.2\\
\hline
\end{tabular}
\end{threeparttable}
\end{table}

\section{Conclusion}\label{ch:conclusion}
This paper proposed a non-imputation-based method for incomplete multimodal learning in MSA: 
Multi-Modal Contrastive Knowledge Distillation (MM-CKD).
We also proposed a novel method to realize effective cross-modal knowledge distillation from the teacher model to student models by MVSC loss. 
Our experiments demonstrated the effectiveness of MM-CKD as a SOTA non-imputation-based method and is competitive with SOTA imputation-based methods. 
While our method only uses about $30\%$ FLOPs of the SOTA imputation-based methods in the test,
it is weak in acoustic and visual feature learning without lexical information, which implies difficulty in transferring high-quality information to auxiliary modalities without strong generative DNNs. 
This point will be the next challenge for future work.

\newpage
% \backmatter
% \include{acknowledgement}
% \newcommand{\BIBdecl}{\setlength{\itemsep}{0.27 em}}
\bibliographystyle{unsrt}
\bibliography{MMM_ref}       % Sample
% \bibliography{main}

\end{document}